\documentclass[12pt,a4paper]{article}
\usepackage{amsmath}
\usepackage{graphicx}
\usepackage{times}
\begin{document}
\begin{titlepage}
\title{Comment on energy dependence of the slope parameter}
\author{ S.M. Troshin, N.E. Tyurin\\[1ex]
\small  \it NRC ``Kurchatov Institute''--IHEP\\
\small  \it Protvino, 142281, Russian Federation}
\normalsize
\date{}
\maketitle

\begin{abstract}
We discuss energy dependence of the slope parameter in elastic proton scattering.
It is shown that unitarity generates energy dependence of the slope parameter in geometrical models   consistent with the experimental results including  recent  LHC data. 

\end{abstract}
\end{titlepage}
\setcounter{page}{2}
\section*{Introduction}
A speeding up of the energy increase of the slope parameter in elastic proton scattering is among  important discoveries performed at the LHC \cite{totem, atlas}. It has been found that the  rate of the slope parameter increase getting larger with  the energy growth compared to the rate at lower energies. It means a speed up  of the interaction radius energy dependence in transverse plane. The reason for this is directly connected to  hadron interaction dynamics in the soft region and makes the studies in that direction important.

It is essential to take into account  that  hadrons are  composite, extended objects and   have formfactors  described by nontrivial functions. Their geometrical radii, contrary to the interaction ones, are energy independent, and  determined by the minimal mass of the  exchanged quanta  responsible for the scattering \cite{yuk}. The generation of the energy--dependent interaction radius is due to unitarity condition in the direct reaction channel. Account of unitarity is performed by the unitarization of an input amplitude.
This  is a way to construct a final scattering amplitude obeying unitarity.

The need for unitarization has become evident at the time when the total cross--sections rise has been discovered. To reconcile the Regge model predictions to the experimental data one should introduce a Pomeron pole  contribution   with the  intercept $\alpha(0)$ greater than unity. Such contribution would finally violate unitarity and therefore requires unitarization. The input amplitude of the Regge model with linear trajectory $\sim (s/s_0)^{\alpha(t)}$, however, includes diffraction cone shrinkage ab initio, i.e. the slope parameter $B(s)$ logarithmically increases with the energy, $B(s)\sim \alpha'(0)\ln (s/s_0)$, $\alpha'(0)\neq 0$, while unitarity requires its double logarithmic asymptotic growth, $B(s)\sim \ln^2(s/s_0)$ if the total cross-section saturates the Froissart-Martin bound. To bring  the slope parameter energy increase with to requirements of unitarity bounds, one can assume  the slope of the Pomeron trajectory $\alpha'(0)$ is an energy--dependent ``effective'' function (cf. \cite{rysk}). 

For the case of the input amplitude assumed by the geometrical models, there is a second reason for unitarization.
The input amplitude itself does not imply growth of slope parameter with energy in geometrical models. 
Only unitarization generates energy dependence of the slope parameter. This parameter grows with energy at its low and moderate values, where total cross--section does not increase. Unitarization makes energy dependence  of the slope parameter consistent with experimental trend at such energies.

In this note  the origin of $B(s)$ growth due to unitarity  is discussed. We consider class  the geometrical models operating with the amplitudes in the impact parameter representation (cf. for definition \cite{heny}). 
 \section{Geometrical models and the slope parameter }
 In these models an input amplitude which is a subject for subsequent unitarization is taken in a factorized form.
  The diffraction cone slope $B^0$ corresponding to such input amplitude does not depend on  energy. It is determined by the geometrical radii of the colliding hadrons in the transverse plane.  The geometrical radius of a hadron  in its turn is determined by a minimal mass of the  exchanged quanta  responsible for the scattering \cite{yuk}. The energy dependence of the actual slope $B(s)$ is generated then through the  unitarization. 
 
 The studies of  geometrical properties of hadron interactions   are important  \cite{blg} for understanding the hadron dynamics ultimately related to the nonperturbative sector of QCD. Under this the unitarity leads to energy dependence of the hadron interaction region in the transverse plane which initially has purely geometrical meaning. Physical interpretation of such mechanism can be found e.g. in \cite{stod}.
 Thus, the final interaction radius appears to be an energy-dependent and so is 
  the quantity 
 \[
 B(s)=\frac {d}{dt} \ln \frac{d\sigma}{dt}|_{t=o}.
 \]

In geometrical  models an input amplitude  is commonly taken as an overlap of the  matter distributions  of the colliding hadrons $D_1  \otimes D_2$ following  the pioneering paper by Chou and Yang \cite{chy}.  It should be noted that the factorization  results also from  the tower diagrams calculations in the electrodynamics \cite{cheng}. 

We  suppose  that the real part of the elastic scattering amplitude is vanishingly small and can be neglected since the high energy experimental data are consistent with the pure 
imaginary amplitude. 
We discuss the slope of the diffraction cone, $B(s)$, which 
is determined by the mean value of the impact parameter squared $b^2$, 
\begin{equation}\label{aver}
\langle b^2\rangle_{tot}=\frac{\int_0^\infty b^3dbf(s,b)}{\int_0^\infty bdbf(s,b)}.
\end{equation}

Account for unitarity is performed by  the schemes which provide an output amplitude $f(s,b)$ limited by the unitarity limit $f=1$  or by the black disc limiting value of $1/2$  \cite{glushko}. 
Mechanisms generating the diffraction cone slope  increase with  energy are similar for the different approaches.
Eq.(\ref{aver}) and the unitarity
\[ 
\mbox{Im} F(s,t)=H_{el}(s,t)+H_{inel}(s,t),
\]
where $H_{el,inel}(s,t)$ are the elastic and inelastic overlap functions, lead to the representation of $B(s)$ as a sum of elastic and inelastic contributions, i.e.
\[
B(s)=B_{el}(s)+B_{inel}(s).
\]Here (cf. \cite{ajduk}), 
\[
B_{el, inel}(s)\sim\frac{\sigma_{el, inel}}{\sigma_{tot}(s)}\langle b^2\rangle_{el, inel}(s).
\]
Thus, the energy dependence of $B(s)$ is determined by the cross-sections $\sigma_{el, inel}$ and average values $\langle b^2\rangle_{el, inel}(s)$. Averaging is going over corresponding overlap functions. Different unitarization schemes provide different asymptotics for $B_{inel}(s)$. 

First, we consider  the unitarization  which incorporates the two scattering modes at high energy:
absorptive and reflective ones and assumes saturation of the unitarity limit by the partial amplitude  at $s\to\infty$ \cite{refl}. We discuss asymptotic energy dependencies of $B_{el, inel}(s)$.
The  $U$--matrix \cite{uma} unitarization scheme incorporates both  modes and   the relation between the scattering amplitude $f$ and the input amplitude $u$ has rational form:
\begin{equation}\label{um}
f(s,b)=u(s,b)/[1+u(s,b)],
\end{equation}
where $u$ is a non-negative function. 

In the geometrical models the  $u(s,b)$ has a factorized form:
\begin{equation}\label{usb}
u(s,b)=g(s)\omega(b),
\end{equation}
where $g(s)\sim s^\lambda$ at  large values of $s$ and $\omega(b)$ exponentially decreases of  at $b\to \infty$ .  The  power dependence on energy guarantees unitarity saturation $f\to 1$ at fixed $b$ and respective asymptotic growth of the total cross--section $\sigma_{tot}\sim \ln^2 s$. Such forms of these functions can be justified by theoretical calculations based on massive quantum electrodynamics \cite{chengwu}. That factorized form and Eq. (\ref{um})  along with the  function $\omega(b)$ chosen to meet the analytical properties of the scattering amplitude lead  to the following asymptotic dependencies
\begin{equation}\label{bes}
B_{el}(s)\sim \ln^2 s
\end{equation}
and
\begin{equation}\label{bis}
B_{inel}(s)\sim \ln s
\end{equation}
since $\langle b^2\rangle_{el, inel}(s)\sim \ln^2 s$ and $\sigma_{el, tot}(s)\sim \ln^2 s$ while 
 $\sigma_{inel}(s)\sim \ln s$ at $\to\infty$. If one would apply those asymptotic dependencies at available energies\footnote{It is a common practice, e.g. Regge model is based on asymptotic dependence of Legendre polynomials, but its results are widely used at modern energies.}, one should then interpret an observed speed up of $B(s)$ growth from  
 $\ln s$ to  $\ln^2 s$ as a transition between the two contributions into the slope. 
 
 In the framework of the geometrical considerations a typical way to construct the function $\omega(b)$ as it was already noted is to represent it as a convolution of the two matter distributions in transverse plane as it was proposed by Chou and Yang \cite{chy}:
\begin{equation}
\omega (b)\sim D_1\otimes D_2\equiv \int D_1({\bf b}_1)D_2({\bf b}-{\bf b}_1).
\end{equation}
This function can also be constructed by taking into account the hadron quark structure \cite{chiral}. 
Thus, the following form was adopted in Eq. (\ref{usb})
\begin{equation}\label{omb}
\omega (b)\sim \exp{(-\mu b)}.
\end{equation}
The value of the energy independent parameter $\mu$ is determined by a particular chosen  model, it can be  assumed that   $\mu=2 m_\pi$ based on the notion of hadrons' peripheral pion cloud.

Thus, the slope parameter $B(s)$ has leading energy dependence at $s\to\infty$ 
\begin{equation}\label{bes}
B(s)\sim \ln^2 s,
\end{equation}
which is related to elastic contribution. Absorption provides a subleading contribution  of Eq. (\ref{bis}).

On the other hand, there is no way to discriminate elastic and inelastic contributions into $B(s)$ at $s\to\infty$ when only the absorptive scattering mode is assumed. In this case both contributions $B_{el}(s)$ and $B_{inel}(s)$ are proportional to $\ln^2 s$ at $s\to\infty$.  It is a typical situation with  $B(s)$ behavior in the unitarization schemes based on eikonal or continued unitarity \cite{glushko}.

\section*{Conclusion}
 Factorization of the input amplitude in geometrical models   can be interpreted as a manifestation of the independence of transverse and longitudinal dynamics in the first approximation. The interrelation of longitudinal and transverse dynamics is then a consequence of the unitarization. The  generation of the $B(s)$ energy growth can be treated due to unitarity alone, namely, the unitarization transforms the input  amplitude with an energy independent  slope into the one with the slope increasing with energy and its increase is proportional to  $\ln^2 s$ at $s\to\infty$.  If the scattering amplitude saturates unitarity limit, the slope parameter $B(s)$ should experience a speed up of its energy dependence at $s\to\infty$.

The most recent experimental data of the TOTEM  and ATLAS-ALFA \cite{totem, atlas} are in favor of transition to the double logarithmic increase of the parameter $B(s)$. The new data are consistent with the proposed mechanism  of the $B(s)$ growth  generation through  unitarity, but the different interpretation cannot be excluded (cf. for a recent review \cite{laslo}) . 

Thus, unitarization  leads  to a nontrivial energy dependence of the slope parameter in the geometric models. It also brings energy dependencies
of $\sigma_{tot}(s)$ and $B(s)$ into agreement with the asymptotic rigorous bounds.

\end{document}